\documentclass[amsmath,amsfonts,prx,twocolumn,longbibliography]{revtex4-1}
\usepackage{graphicx}
\usepackage{epsfig}
\usepackage{bbm}
\usepackage{bm}
\usepackage{dcolumn}
\usepackage{amsmath}
\usepackage{amssymb}
\usepackage{color}
\usepackage[varg]{txfonts}

\newcommand{\beg}{\begin{equation}}
\newcommand{\en}{\end{equation}}

\newcommand{\bk}{\mathbf k}
\newcommand{\br}{\mathbf r}

\newcommand{\bn}{\mathbf n}

\newcommand \bel  {\begin{align}}
\newcommand \enl  {\end{align}}

\newcommand{\veps}{\varepsilon}

\begin{document}

\title{Inverse Faraday effect in superconductors with potential impurities}

\author{Maxim Dzero}
\affiliation{Department of Physics, Kent State University, Kent, OH 44242, USA}

\begin{abstract}
I consider a nonlinear response of conventional superconductors contaminated with nonmagnetic impurities or imperfections to a circular polarized light. I focus on dc contributions to the induced current density which describe the emergence of the static magnetization in a superconductor. This effect is known as inverse Faraday effect. By employing quasiclassical theory of superconductivity I derive an expression for the induced static magnetization as a function of frequency of external ac field and disorder scattering rate. The scattering of electrons off potential impurities is taken into account within the framework of the self-consistent Born approximation. It is found that the magnitude of the inverse Faraday effect decreases with an increase in disorder scattering rate.
I have also discovered that the value of the induced magnetization has a characteristic minimum at a frequency which approximately equals twice the value of the pairing gap in a clean superconductor. This minimum appears due to the resonant excitation of the amplitude Higgs mode.  
\end{abstract}

\date{\today}

\maketitle

\section{Introduction}
In the last decade there have been significant advances in state-of-the-art optical measurements and techniques 
that have spurred a lot of interest in the physical phenomena related to
nonlinear responses in conventional and unconventional superconductors \cite{THz1,THz2,Shimano2012,Shimano2013,Shimano2014,THz3,Millis2020,Armi2014,Armitage2023}. These experimental and theoretical studies usually uncover and address a number of interesting and important problems. 
Some of these problems are concerned with the fundamental aspects of superconductivity such as various proposals to probe the physics associated with the excitation the amplitude Higgs mode \cite{Varma2014,Cea2016,Shimano2020,Haenel2021-Disorder,Sherman2015-Disorder,Spivak-Higgs} and Eliashberg effect \cite{Eliashberg1970,Eremin2023} to name a few. Another set of problems focuses on development of potential applications of various nonlinear effects in superconductor-based devices for it offers one a possibility to manipulate the physical properties of superconductors such as magnetic and transport response functions on extremely short time scales \cite{Chen2017,Croitoru2022,SpivakGiant2020}.
 
\begin{figure}
\includegraphics[width=0.835\linewidth]{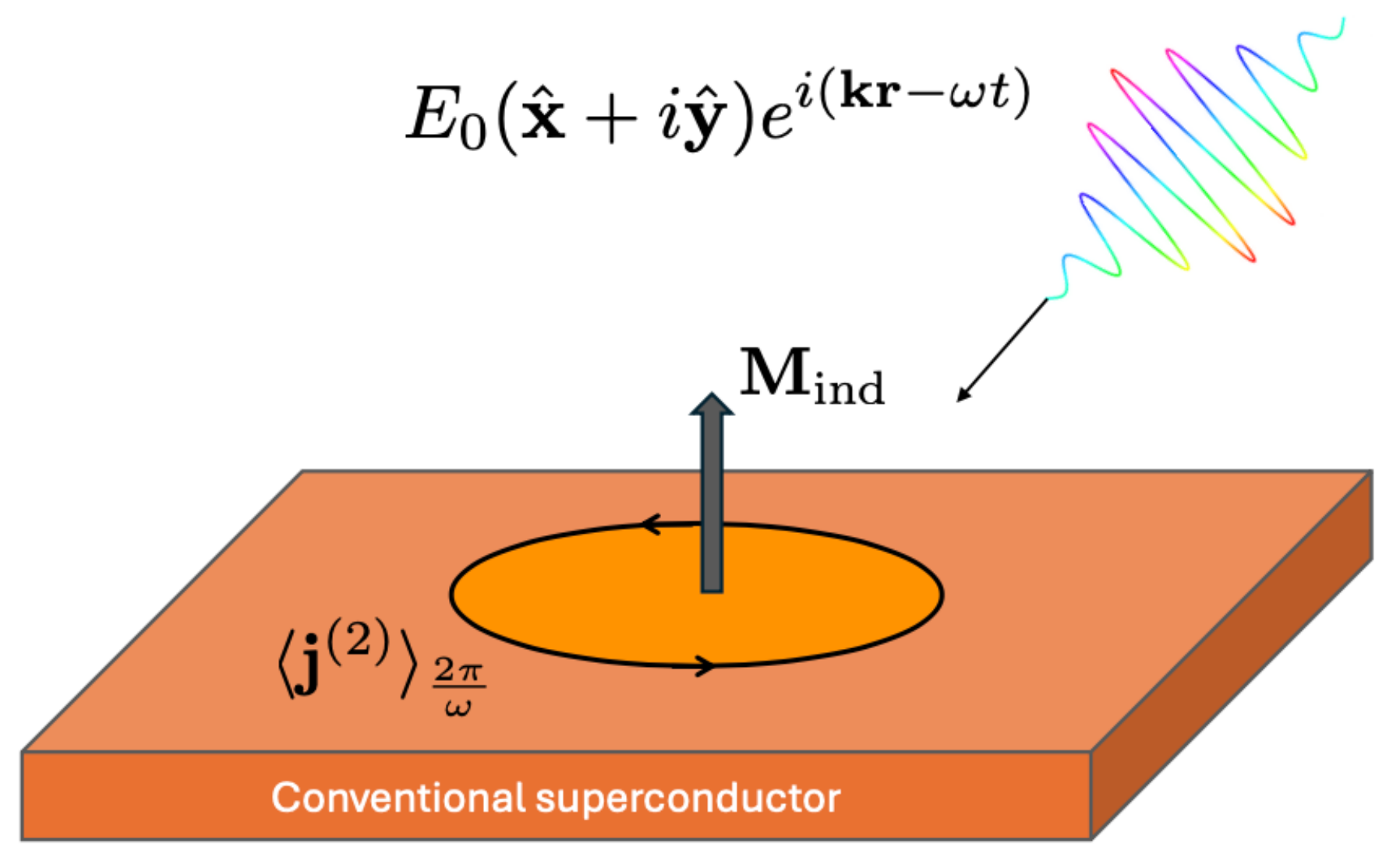}
\caption{Schematic representation for the realization of inverse Faraday effect in superconductors. When a conventional superconductor is subjected to a circular polarized light with frequency $\omega$, it induces a static component in the current density and, as a consequence, a static magnetization ${\mathbf M}_{\textrm{ind}}$. The effect is of the second order in the powers of the amplitude $E_0$ of an external electric field.}
\label{Fig1Main}
\end{figure}

Inverse Faraday effect (IFE) in dispersive medium was discovered by L. P. Pitaevskii \cite{Pit1961} who demonstrated that static magnetization can be produced by subjecting a medium to high-intensity circular polarized light. In subsequent years this effect has been and continues to be actively studied in various electronic systems \cite{Pershan1966,Battiato2014,yang2022inverse,yang2022inverse,mou2023reversed,mou2023chiral,IFE-Mott2022,gao2020topological,parchenko2023plasmonenhanced,Han_2023}.
For example, Majedi \cite{Majedi2021}, using phenomenological set of arguments, has recently proposed that the IFE can be induced in superconductors by subjecting them to an external microwave radiation, Fig. \ref{Fig1Main}. Almost immediately after the publication of that work, several groups have further developed this idea by studying the IFE in superfluid condensates \cite{Mironov2021-IFESC} and in various superconductor-based devices \cite{Parafilo2022Fl,Putilov2023-IFESC,Buzdin2023,Croitoru2023}. Most recently, Sharma and Balatsky \cite{Balatsky2023} has utilized the quasiclassical approach to formulate a microscopic theory of the IFE in superconductors. 

In the context of the present work it must be mentioned that in the earlier works \cite{Majedi2021,Balatsky2023} scattering due to potential disorder, which is almost inevitably present in superconducting samples, has been taken into account on purely phenomenological level by formally assigning finite imaginary part to the single particle energies, $\veps\to\veps+i\Gamma$. It is well known that weak potential disorder does not affect the physical properties of conventional superconductors in equilibrium (Anderson theorem) \cite{AndersonTheorem,AG1961,Balatsky-RMP} and for that reason the approximations taken in the above mentioned works may seem well justified. However, as it has been established independently by several groups, potential impurities do affect the superconducting response functions provided that a superconductor has been driven out-of-equilibrium  \cite{Silaev2019-Disorder,Yang2022-Disorder,Yang2022-Disorder2,Dzero2023-Disorder}.

Within the theoretical framework of the quasiclassical theory of superconductivity \cite{LO} this can be understood as follows. The self-energy due to scattering on potential impurities $\check{\Sigma}$ enters in the quasiclassical equations in a commutator with the single-particle matrix correlation function $\check{g}$. Since in Nambu representation the self-energy part due to disorder scattering is proportional to $\check{g}$ the corresponding commutator vanishes identically. Qualitatively, this means that Cooper pairs scatter coherently off the weak potential impurities and pair breaking processes do not occur. When a superconductor is driven out of equilibrium, the commutator will include the derivatives of both $\check{\Sigma}$ and $\check{g}$ with respect to time and, as a result, no such cancellation occurs. In other words, the monochromatic radiation breaks time-reversal symmetry and, as a result, scattering on potential impurities induces pair-breaking. Therefore, in the context of the IFE in superconductors, it is important to elucidate the role of nonmagnetic disorder and how it affects the ac-field induced dc-magnetization. 

In this paper I formulate a microscopic theory of inverse Faraday effect in disordered superconductors in which the effects of potential disorder are treated self-consistently. In particular I demonstrate that under the action of external ac electric field the self-energy correction due to disorder scattering leads to the renormalization of the quasiparticle energies and pairing amplitude similar to the renormalization effects which appear due to the presence of paramagnetic impurities. I analyze the IFE by calculating the static contribution to the charge current density to the second order in external electric field. I have found that the magnitude of the induced dc magnetization decreases with the increase in disorder scattering rate. Intriguingly, I also find that the magnetization as a function of disorder scattering rate is suppressed most strongly when the frequency of electric field coincides with the resonant frequency of the amplitude Higgs mode $\omega_{\textrm{res}}$. Furthermore, magnetization as a function of frequency for a fixed disorder strength exhibits a minimum at $\omega=\omega_{\textrm{res}}$. Therefore, based on these observations I suggest that induced dc magnetization via IFE in conventional superconductors contains a 'fingerprint' of the amplitude Higgs mode. 
\section{Qualitative discussion}
When a conventional superconductor is subjected to a monochromatic external electric field
\beg\label{Field}
{\mathbf E}_{\textrm{ext}}(\br,t)={\mathbf E}_0e^{i\bk \br-i\omega t}+{\mathbf E}_0^*e^{-i\bk \br+i\omega t}
\en
with ${\mathbf E}_0=E_0(\hat{\mathbf x}+i\hat{\mathbf y})$ (here $\hat{\mathbf x}$ and $\hat{\mathbf y}$ are the unit vectors), Fig. \ref{Fig1Main},
the superconducting order parameter, defined by a scalar quantity, must couple to the electric field nonlinearly. As a result of this nonlinear coupling two collective modes will be excited: Carlson-Goldman (phase) mode and amplitude Higgs mode. In the context of neutral superfluids, the former is referred to as Anderson-Bogoliubov mode \cite{Kamenev2011}. Usually, these modes are analyzed separately because in the Nambu representation they are described by the correlation functions with different matrix structure (see discussion in the Section III below). In addition, the phase mode is gapless while the amplitude mode is gapped at twice the value of the pairing amplitude in equilibrium, 
$\omega_{\textrm{res}}=2\Delta_0$. 

As it turns out there are several physical effects, such as inverse Faraday and ponderomotive force effects, for which an interplay between phase and amplitude modes becomes crucial and so it could be important for deeper understanding of these effects to find out how this interplay can be identified in the physical observables. In order to succeed in achieving this goal, one needs to treat these modes on equal footing. For example, in the IFE the phase fluctuations produce the correction to the particle density distribution which, due to the fact that phase mode is gapless, must be linearly proportional to the gradient of an external electric field, $\delta n\sim ({\mathbf v}_F\cdot{\mbox{\boldmath $\nabla$}})(\bn\cdot{\mathbf E}_{\textrm{ext}}(\br,t))$ (here ${\mathbf v}_F=v_F\bn$ is the Fermi velocity). On the other hand, the amplitude mode is gapped and, as a result, fluctuations associated with it produce the correction to particle velocity $\delta{\mathbf v}\sim {\mathbf v}_F(\bn\cdot{\mathbf E}_{\textrm{ext}}(\br,t))$ and gradient corrections can only appear in the second order, i.e. the amplitude mode is diffusive, $\omega_{\textrm{amp}}(q)\approx2\Delta_0+Dq^2$ (here $D=v_F^2\tau/2$ is the diffusion coefficient for a two-dimensional superconductor).  As a result, the second order correction to the current in powers of electric field can be computed using the hydrodynamic expression ${\mathbf j}^{(2)}=\langle\langle\delta n\delta {\mathbf v}\rangle_{\bn}\rangle_{2\pi/\omega}$, where the averaging is performed over the directions of the Fermi velocity and period of oscillating electric field. Among several contributions, an expression for ${\mathbf j}^{(2)}$ will contain the terms describing the gyration of electric field which gives rise to the \emph{static magnetization} ${\mathbf M}_{\textrm{ind}}(\omega)\propto{\mathbf E}_{\textrm{ext}}\times{\mathbf E}_{\textrm{ext}}^*$ (see Fig. \ref{Fig1Main}). From quite general considerations one may immediately discover that ${\mathbf M}_{\textrm{ind}}(\omega)\propto e^2v_F^4/\omega^3$ and one can represent the resulting expression for the induced magnetization as ${\mathbf M}_{\textrm{ind}}(\omega)=ie^2v_F^4\gamma_\omega({\mathbf E}_{\textrm{ext}}\times{\mathbf E}_{\textrm{ext}}^*)$,
where function $\gamma_\omega\propto 1/\omega^3$ is determined by the microscopic details such as disorder, interactions between the constituent quasiparticles etc. \cite{Balatsky2023} 

The excitation of the phase mode corresponds to the re-distribution of the electronic charge for the fluctuations in the phase of the superconducting order parameter are conjugate to the fluctuations in the particle number by virtue of the Heisenberg uncertainty relation $\delta N\delta\phi\sim\hbar$ \cite{Kamenev2011}. In the ground state and in the absence of the supercurrent, the superconducting order parameter can always be chosen to be real which corresponds to the particle-hole symmetric case. Then, under the action of an external ac field the phase mode will be excited leading to a particle-hole asymmetry. Therefore one immediately concludes that the emergence of the IFE in superconductors requires particle-hole asymmetry. 

In the context of IFE in superconductors this aspect of the problem - importance of the particle-hole symmetry - has been previously emphasized by several groups \cite{Mironov2021-IFESC,Putilov2023-IFESC,Buzdin2023,Balatsky2023}. Specifically, Mironov et al. \cite{Mironov2021-IFESC} have studied how superconducting condensates may acquire a temperature dependent dc magnetic moment under the action of the external circularly polarized light by considering the phenomenological model based on the time-dependent Ginzburg-Landau theory. Even though the physics associated with phase fluctuations of the order parameter has been properly addressed, the model used in Ref. \cite{Mironov2021-IFESC} has a number of important shortcomings such as a quite restricted range of validity (i.e. temperatures must be close to the critical temperature) and an assumption of the gapless superconductivity. This latter assumption does not allow one to properly account for the contributions from the excitation of the amplitude mode to the IFE. 

Indeed, the fact that the excitation of the amplitude mode is also crucial for the IFE in superconductors has been surprisingly overlooked so far. Furthermore, given the discussion above it follows that the induced magnetization must also exhibit a non-monotonic frequency dependence - a minimum -  around $\omega\approx 2\Delta_0$ which corresponds to the resonant frequency for the excitation of the amplitude Higgs mode (see above). However, in order to check how prominent this feature will be it requires a microscopic calculation. 
In passing I note that since the amplitude fluctuations can only produce second order gradient corrections, these fluctuations alone cannot contribute to the IFE and so at this point it is not immediately clear how to capture this effect if one chooses to work in the dirty limit $\tau\Delta_0\ll 1$ by employing the Usadel equation \cite{Usadel1970}. In the following Section I provide the microscopic derivation of the expression for the function $\gamma_\omega$ using the quasiclassical theory of superconductivity \cite{LO,Usadel1970,Kita2001}.

\section{Main equations}
In the problems which are concerned with response of superconductors to the application of external electromagnetic radiation, it proves important to write down the equations for the gauge invariant single-particle propagators \cite{Kopnin1994,Kita2001}
\beg\label{checkg}
\check{g}(1,2)=\left(\begin{matrix} \hat{g}^R(1,2) & \hat{g}^K(1,2) \\ 0 & \hat{g}^A(1,2)\end{matrix}\right)
\en
within the framework of the quantum field theory for non-equilibrium systems \cite{Kamenev2009,Kamenev2011}.
Each component of the matrix function $\check{g}(1,2)$ is defined in Keldysh and Nambu spaces and the argument of the propagator should be understood as $(1,2)=(\br_1,t_1;\br_2,t_2)$. 
One can follow the avenue of Refs. \cite{Kita2001,Kita-Review} to derive the quasiclassical equation for the Wigner transformed single-particle propagator $\check{g}$:
\beg\label{EilenFin}
\begin{split}
&[\veps\check{\tau}_3,\check{g}]+\frac{e}{c}{\mathbf v}_F{\mathbf A}(\br,t)[\check{\tau}_3,\check{g}]+i{\mathbf v}_F{\partial_\br}\check{g}-[\check\Sigma\check{\tau}_3\stackrel{\circ},\check{g}]
+\frac{ie}{2}\\&\times\left[\partial_\veps\left(\check{g}\check{\tau}_3\right){\cal E}\left(i\stackrel{\leftrightarrow}{\partial}_{\veps t}\right)
+\partial_\veps\left(\check{\tau}_3\check{g}\right){\cal E}\left(-i\stackrel{\leftrightarrow}{\partial}_{\veps t}\right)\right]{\mathbf v}_F{\mathbf E}(\br,t)=0.
\end{split}
\en
In this equation $\stackrel{\leftrightarrow}{\partial}_{\veps t}=\frac{1}{2}\stackrel{\leftarrow}\partial_\veps\stackrel{\rightarrow}\partial_t$, $\check{g}=\check{g}(\bn\veps;\br t)$, $\br=(\br_1+\br_2)/2$, $t=(t_1+t_2)/2$, ${\mathbf v}_F=v_F\bn$ is the Fermi velocity, ${\mathbf A}(\br,t)$ is a vector potential, ${\mathbf E}(\br,t)=-(1/c)\partial_t{\mathbf A}(\br,t)$ is an external electric field, $\check{\tau}_3$ is the third Pauli matrix which is diagonal in Keldysh subspace, function ${\cal E}(u)=(e^u-1)/u$, $\check{\Sigma}$ is the self-energy part which encodes the effects of superconducting pairing and disorder:
\beg\label{SelfEnergy}
\check{\Sigma}\check{\tau}_3=-\frac{i}{2\tau}\langle \check{g}\rangle_\bn+i\check\tau_2\Delta'+i\check{\tau}_1\Delta'',
\en
$\check{\tau}_{1(2)}=\hat{\mathbbm{1}}_{2\times2}\otimes\hat{\tau}_{1(2)}$ are first and second Pauli matrices, which act in Nambu and Keldysh spaces and are diagonal in Keldysh space. In all the mathematical expressions throughout the text which include two vectors ${\mathbf a}$ and ${\mathbf b}$ in the combination ${\mathbf a}{\mathbf b}$ the dot-product is always implied.
Lastly, the commutator is defined according to
\beg\label{CommDefine}
[A\stackrel{\circ},B]=Ae^{\frac{i}{2}\left(\stackrel{\leftarrow}\partial_{\veps}\stackrel{\rightarrow}\partial_t-\stackrel{\leftarrow}\partial_{t}\stackrel{\rightarrow}\partial_\veps\right)}B-Be^{\frac{i}{2}\left(\stackrel{\leftarrow}\partial_{\veps}\stackrel{\rightarrow}\partial_t-\stackrel{\leftarrow}\partial_{t}\stackrel{\rightarrow}\partial_\veps\right)}A.
\en
Given the expression for the electric field (\ref{Field}), I will look for the solution of (\ref{EilenFin}) in the form
\beg\label{Four}
\begin{split}
\check{g}(\bn\veps;\br t)=\check{g}(\bn\veps;\bk\omega)e^{i(\bk\br-\omega t)}+\check{g}(\bn\veps;-\bk,-\omega)e^{-i(\bk\br-\omega t)}.
\end{split}
\en
It is important to emphasize here that equation (\ref{EilenFin}) does not preserve the norm of $\check{g}$, i.e. $\check{g}^2=\hat{\mathbbm{1}}$ will be satisfied only when either ${\mathbf E}(\br,t)=0$ or in the limit $|\veps|\to \infty$ \cite{Kita2001}. The similar situation happens in a problem of non-equilibrium dynamics of the a superconductor with paramagnetic impurities \cite{Dzero2023-Disorder} which has one common feature with our problem at hand: the lack of time-reversal symmetry. 

In the ground state, the Keldysh propagator is just a parametrization 
\beg\label{gK}
\hat{g}_\veps^K=(\hat{g}_\veps^R-\hat{g}_\veps^A)\tanh\left(\frac{\veps}{2T}\right),
\en
where $T$ is temperature. In the ground state I can set $\Delta'=\Delta_0$, $\Delta''=0$ and choose $\hat{g}_\veps^{R(A)}=g_{\veps}^{R(A)}\hat{\tau}_3-if_{\veps}^{R(A)}\hat{\tau}_2$. By employing the normalization condition $\hat{g}_\veps^{R(A)}\hat{g}_\veps^{R(A)}=\hat{\mathbbm{1}}$ I find $g_\veps^{R(A)}=(\veps\pm i0)/\zeta_\veps^{R(A)}$, $f_\veps^{R(A)}=\Delta_0/\zeta_\veps^{R(A)}$ and
\beg\label{zetaveps}
\zeta_\veps^{R(A)}=\left\{\begin{matrix}\pm\textrm{sign}(\veps) \sqrt{(\veps\pm i0)^2-\Delta_0^2}, \quad |\veps|>\Delta_0, \\
i\sqrt{\Delta_0^2-\veps^2}, \quad |\veps|<\Delta_0.
\end{matrix}\right.
\en
The value of the superconducting order parameter in equilibrium is then determined self-consistently from
\beg\label{Self}
\Delta_0=\frac{\lambda}{2}\int\limits_{-\omega_D}^{\omega_D} f_\veps^Kd\veps,
\en
where $\lambda$ is the dimensionless interaction strength and $\omega_D$ is the Debye frequency. 
Naturally, the disorder scattering rate does not enter into this expressions since the commutator in (\ref{EilenFin}) with the first term in (\ref{SelfEnergy}) vanishes in equilibrium. 

The expressions for the particle density and particle velocity for a given ${\mathbf v}_F$ are expressed in terms of the Keldysh propagator as follows:
\beg\label{denvel}
\begin{split}
n(\bn;\br t)&=\frac{\pi\nu_F}{2}\int\textrm{Tr}\left\{\hat{g}^K(\bn\veps;\br t)\right\}d\veps, \\
{\mathbf v}(\bn;\br t)&=\frac{\pi\nu_F}{2}\int{\mathbf v}_F\textrm{Tr}\left\{\hat{\tau}_3\hat{g}^K(\bn\veps;\br t)\right\}d\veps.
\end{split}
\en
Here $\nu_F$ is the single particle density of states at the Fermi level.
In terms of these quantities, the macroscopic current density is defined as
\beg\label{CurrentDensity}
{\mathbf j}(\br,t)=\langle n(\bn;\br,t){\mathbf v}(\bn;\br,t)\rangle_\bn.
\en
Although the expressions above are valid in any spacial dimenions, for the calculation of the averages over the Fermi surface, it will be assumed without loss of generality that I have a two-dimensional superconductor, Fig. \ref{Fig1Main}. 
As I have discussed above, our goal is to compute linear corrections for the Keldysh Green's function $\hat{g}^K$ in electric field, which will allow us to evaluate nonlinear in electric field correction to the current density (\ref{CurrentDensity}). 

\subsection{Linear analysis: retarded and advanced functions}
I start by computing the linear corrections to $\hat{g}^{R(A)}$. I will look for the solution of equation (\ref{EilenFin}) with 
\beg\label{LinCorr}
\hat{g}^{R(A)}=\hat{g}_\veps^{R(A)}+\hat{g}_1^{R(A)},
\en
where $\hat{g}_1^{R(A)}\propto \bn\cdot{\mathbf E}_0$.
Inserting this expression into (\ref{EilenFin}) and keeping only terms to the linear order in ${\mathbf E}_0$ I obtain the following equation for the matrix function $\hat{g}_1^{R(A)}(\bn\veps;\bk\omega)$:
\beg\label{RALinear}
\begin{split}
&[\veps\hat{\tau}_3,\hat{g}_1]+\omega\{\hat{\tau}_3,\hat{g}_1\}-v_F(\bn\bk)\hat{g}_1-i\Delta_0[\hat{\tau}_2,\hat{g}_1]\\&-\frac{i}{2\tau}\left(\hat{g}_1-\langle \hat{g}_1\rangle_\bn\right)\hat{g}_{\veps-\frac{\omega}{2}}+
\frac{i}{2\tau}\hat{g}_{\veps+\frac{\omega}{2}}\left(\hat{g}_1-\langle\hat{g}_1\rangle_\bn\right)\\&=
\frac{\bn{\mathbf Q}_0}{\omega}\{\hat{\tau}_3,\hat{g}_{\veps-\omega/2}-\hat{g}_{\veps+\omega/2}\}+i\delta\overline{\Delta}_{\bk\omega}\hat{\tau}_1\hat{g}_{\veps-\omega/2}\\&-i\delta\overline{\Delta}_{\bk\omega}\hat{g}_{\veps+\omega/2}\hat{\tau}_1+
\frac{\bn{\mathbf Q}_0}{\omega}[\hat{\tau}_3,\hat{g}_{\veps-\omega/2}+\hat{g}_{\veps+\omega/2}],
\end{split}
\en
where ${\mathbf Q}_0=iev_F{\mathbf E}_0$, function $\delta\overline{\Delta}_{\bk\omega}$ is the Fourier component of the linear correction to the superconducting order parameter $\delta\Delta(\br,t)=\delta\overline{\Delta}_{\bk\omega} e^{i(\bk\br-\omega t)}$ and I suppressed the superscripts $R(A)$ for brevity. In deriving this equation I have used the auxiliary expressions listed in Appendix A. 

Let me first average both sides of this equation over the Fermi surface. Denoting $\langle\hat{g}_1(\bn\veps;\bk\omega)\rangle_\bn=\hat{\cal J}_{\bk\omega}(\veps)$ I have
\beg\label{RALinearAv}
\begin{split}
&[\hat{\veps},\hat{\cal J}_{\bk\omega}]+\omega\{\hat{\tau}_3,\hat{\cal J}_{\bk\omega}\}=i\delta\overline{\Delta}_{\bk\omega}[\hat{\tau}_1,\hat{g}_\veps]
+v_F\langle\bn\bk \hat{g}_1\rangle_\bn,
\end{split}
\en
where a new notation $\hat{\veps}=\veps\hat{\tau}_3-i\Delta_0\hat{\tau}_2$ has been introduced for brevity. Two comments are in order: (1) note that the terms which contain disorder scattering rate do not appear in this equation; (2) here I have neglected the higher order derivatives of the bare propagator by replacing   the commutator $[\delta\hat{\Delta}(\br,t)\stackrel{\circ},\hat{g}]$ with its equilibrium expression $[\delta\hat{\Delta}(\br,t),\hat{g}]$ since the terms which contain $\delta\Delta_{\bk\omega}$ will be irrelevant for the IFE. Because $\delta\overline{\Delta}_{\bk\omega}$ must be proportional to the electric field, it will be convenient to use function $\delta\Delta_{\omega}$ instead, which is defined by $\delta\overline{\Delta}_{\bk\omega}=(\bk{\mathbf Q}_0/2)\delta\Delta_{\omega}$. By the same token I will re-define $\hat{\cal J}_{\bk\omega}(\veps)=(\bk{\mathbf Q}_0/2)\delta\hat{\cal G}_\veps(\omega)$. In what follows I can also neglect the anomalous part of the third term on the right hand side of equation (\ref{RALinearAv}) since $v_Fk=(v_F/c)\omega\ll \omega$. The same approximation has already been used in deriving equation (\ref{EilenFin}) by ignoring the contributions from the magnetic field \cite{Balatsky2023,Kita2001}.  

Next, I multiply both sides of equation (\ref{RALinear}) by $\bn\cdot\bk$ and average the resulting expression over the Fermi surface. This yields:
\beg\label{Eq4g1RA}
\begin{split}
&\left(\tilde{\veps}_{\omega}^{R(A)}\hat{\tau}_3-i\tilde{\Delta}_{\omega}^{R(A)}\hat{\tau}_2\right)\hat{\cal G}_\veps^{R(A)}-\hat{\cal G}_\veps^{R(A)}
\left(\tilde{\veps}_{-\omega}^{R(A)}\hat{\tau}_3-i\tilde{\Delta}_{-\omega}^{R(A)}\hat{\tau}_2\right)\\&=a_\veps^{R(A)}(\omega)\hat{\tau}_0-b_\veps^{R(A)}(\omega)\hat{\tau}_1,
\end{split}
\en 
where $\langle(\bn\bk)\hat{g}_1(\bn\veps;\bk\omega)\rangle_\bn=({\bk{\mathbf Q}_0}/2)\hat{\cal G}_\veps^{R(A)}(\omega)$ and I have introduced functions 
\beg\label{Shifted}
\begin{split}
\tilde{\veps}_\omega^{R(A)}&=\omega+\veps+\frac{i}{2\tau}g_{\veps+\frac{\omega}{2}}^{R(A)}, \\
\tilde{\Delta}_\omega^{R(A)}&=\Delta_0+\frac{i}{2\tau}f_{\veps+\frac{\omega}{2}}^{R(A)}
\end{split}
\en
and $a_\veps^{R(A)}(\omega)=(g_{\veps-\omega/2}^{R(A)}-g_{\veps+\omega/2}^{R(A)})/\omega$, $b_\veps^{R(A)}(\omega)=(f_{\veps-\omega/2}^{R(A)}+f_{\veps+\omega/2}^{R(A)})/\omega$. For $\omega=0$ formulas (\ref{Shifted}) coincide with the corresponding expressions in \cite{AG1961}. By analyzing the matrix combinations which enter into both equations (\ref{RALinearAv},\ref{Eq4g1RA}), it is straightforward to realize that both of these equations can be solved using the following ansatz 
\beg\label{GRA}
\begin{split}
&\hat{\cal G}_\veps^{R(A)}(\omega)={\cal G}_\veps^{R(A)}(\omega)\hat{\tau}_3-i{\cal F}_\veps^{R(A)}(\omega)\hat{\tau}_2, \\
&\delta\hat{\cal G}_\veps^{R(A)}(\omega)=\delta{\cal G}_\veps^{R(A)}(\omega)\hat{\tau}_0+\delta {\cal F}_{\veps}^{R(A)}(\omega)\hat{\tau}_1. 
\end{split}
\en
To summarize, the expression for the linear correction to the retarded and advanced parts of $\check{g}$ is formally given by
\beg\label{g1Ansatz}
\hat{g}_{1}^{R(A)}(\bn\veps;\bk \omega)=\hat{\cal G}_\veps^{R(A)}(\omega)\bn{\mathbf Q}_0+\delta\hat{\cal G}_\veps^{R(A)}(\omega)(\bn\bk)(\bn{\mathbf Q}_0).
\en
Expressions for the components of the functions $\hat{\cal G}_\veps^{R(A)}(\omega)$ are listed in Appendix B. Components of $\delta\hat{\cal G}_\veps^{R(A)}(\omega)$ can be easily found:
\beg\label{deltaGRA}
\delta{\cal G}_\veps^{R(A)}(\omega)=
\frac{v_F}{2\omega}{\cal G}_\veps^{R(A)}(\omega), \quad \delta {\cal F}_{\veps}^{R(A)}(\omega)=-\frac{\delta\Delta_\omega}{\zeta_\veps^{R(A)}}. 
\en
Note that there are contributions to $\delta{\cal G}_\veps^{R(A)}(\omega)$ which cancel out and this is the reason why I have had to retain the normal component of $\langle \bn\bk\hat{g}_1(\bn\veps;\bk\omega)\rangle_\bn$ in equation (\ref{RALinearAv}). Furthermore, as I will show below this correction plays a crucial role in determining the magnitude of the IFE. 

\subsection{Linear analysis: Keldysh function}
The Keldysh components of the quasiclassical function $\check{g}_1$ can be computed in full analogy with the calculation of $\hat{g}_1^{R(A)}$. In particular, since the equation for $\langle \bn\bk\hat{g}_1^{K}\rangle_\bn$ is identical to (\ref{RALinearAv}) for the normal component of $\delta\hat{\cal G}_\veps^K(\omega)$ I have
\beg\label{deltaGK}
\delta{\cal G}_\veps^{K}(\omega)=
\frac{v_F}{2\omega}{\cal G}_\veps^{K}(\omega).
\en
I proceed with the equation for the Keldysh component of $\hat{\cal G}_\veps^K(\omega)$ which reads:
\beg\label{g1Keq}
\begin{split}
&\left(\tilde{\veps}_{\omega}^R\hat{\tau}_3-i\tilde{\Delta}_\omega^R\hat{\tau}_2\right)\hat{\cal G}_\veps^K-\hat{\cal G}_\veps^K\left(\tilde{\veps}_{-\omega}^A\hat{\tau}_3
-i\tilde{\Delta}_{-\omega}^A\hat{\tau}_2\right)\\&=\tilde{a}_{\veps}^K(\omega)\hat{\tau}_0-\tilde{b}_\veps^K(\omega)\hat{\tau}_1.
\end{split}
\en
This equation is, of course, quite similar to the equation (\ref{Eq4g1RA}) for the retarded and advanced functions above.
Functions appearing in the right hand side of this equation are defined as:
\beg\label{Coeffs}
\begin{split}
\tilde{a}_{\veps}^K(\omega)=a_\veps^K(\omega)+\frac{i}{2\tau}&\left({\cal G}_\veps^Rg_{\veps-\frac{\omega}{2}}^K-g_{\veps+\frac{\omega}{2}}^K{\cal G}_\veps^A\right.\\&\left.
+f_{\veps+\frac{\omega}{2}}^K{\cal F}_\veps^A-{\cal F}_\veps^Rf_{\veps-\frac{\omega}{2}}^K\right), \\
\tilde{b}_{\veps}^K(\omega)=b_\veps^K(\omega)+\frac{i}{2\tau}&\left(f_{\veps+\frac{\omega}{2}}^K{\cal G}_\veps^A-g_{\veps+\frac{\omega}{2}}^K{\cal F}_\veps^A\right.\\&\left.+{\cal G}_\veps^Rf_{\veps-\frac{\omega}{2}}^K-{\cal F}_\veps^Rg_{\veps-\frac{\omega}{2}}^K\right),
\end{split}
\en
where $a_\veps^{K}(\omega)=(g_{\veps-\omega/2}^{K}-g_{\veps+\omega/2}^{K})/\omega$ and $b_\veps^{K}(\omega)=(f_{\veps-\omega/2}^{K}+f_{\veps+\omega/2}^{K})/\omega$. In the case of strong disorder when $\tau\Delta_0\ll 1$, I see that both functions (\ref{Coeffs}) are linearly dependent on $\tau^{-1}$.
I should emphasize here that no such terms $\propto\tau^{-1}$ will appear in the expression for the Keldysh function if one chooses to work in the dirty limit from the outset by using the Usadel equation \cite{Moore2017,Eremin2023,Yantao2023}. 

The solution of this matrix equation can be found in the form $\hat{\cal G}_\veps^{K}(\omega)={\cal G}_\veps^{K}(\omega)\hat{\tau}_3-i{\cal F}_\veps^{K}(\omega)\hat{\tau}_2$ with
\beg\label{FirstOrderSolK}
\begin{split}
{\cal G}_\veps^K(\omega)&=\frac{\tilde{a}_\veps^K(\omega)(\tilde{\veps}_{\omega}^R+\tilde{\veps}_{-\omega}^A)+\tilde{b}_\veps^K(\omega)(\tilde{\Delta}_{\omega}^R-\tilde{\Delta}_{-\omega}^A)}{[\tilde{\veps}_{\omega}^R]^2-[\tilde{\veps}_{-\omega}^A]^2-[\tilde{\Delta}_{\omega}^R]^2+[\tilde{\Delta}_{-\omega}^A]^2}, \\
{\cal F}_\veps^K(\omega)&=\frac{\tilde{b}_\veps^K(\omega)(\tilde{\veps}_{\omega}^R-\tilde{\veps}_{-\omega}^A)+\tilde{a}_\veps^K(\omega)(\tilde{\Delta}_{\omega}^R+\tilde{\Delta}_{-\omega}^A)}{[\tilde{\veps}_{\omega}^R]^2-[\tilde{\veps}_{-\omega}^A]^2-[\tilde{\Delta}_{\omega}^R]^2+[\tilde{\Delta}_{-\omega}^A]^2}.
\end{split}
\en
I am providing the expression for ${\cal F}_\veps^K(\omega)$ for completeness since this function does not contribute to the IFE. I note that temperature enters into these expressions only through the relation (\ref{gK}) and the temperature dependent value of the pairing gap $\Delta_0(T)$. 
\subsection{Induced static magnetization}
Having computed the linear-in-electric-field corrections to the Keldysh Green's function (\ref{deltaGK},\ref{FirstOrderSolK}) I go back to the hydrodynamic expression for the current density (\ref{CurrentDensity}). Since linear correction to both particle density and velocity will be determined by the normal components of $\check{g}_1^K$, it will be convenient to represent the normal component of $\hat{g}_1^K$ as ${g}_1^K(\bn\veps;\br t)={\cal G}_1^K(\bn\veps;\br t)+\delta{\cal G}_1^K(\bn\veps;\br t)$ with
\beg\label{GKsrt}
\begin{split}
{\cal G}_1^K(\bn\veps;\br t)&=iev_F{\cal G}_\veps^K(\omega)\bn{\mathbf E}(\br,t)\\&+iev_F
{\cal G}_\veps^K(-\omega)\bn{\mathbf E}^*(\br,t)
\end{split}
\en
and
\beg\label{GKsrtp}
\begin{split}
\delta{\cal G}_1^K(\bn\veps;\br t)&=\frac{ev_F^2}{2\omega}{\cal G}_\veps^{K}(\omega)(\bn{\mbox{\boldmath $\nabla$}})\bn{\mathbf E}(\br,t)
\\&-\frac{ev_F^2}{2\omega}{\cal G}_\veps^{K}(-\omega)(\bn{\mbox{\boldmath $\nabla$}})\bn{\mathbf E}^*(\br,t).
\end{split}
\en
Here the electric field ${\mathbf E}(\br,t)={\mathbf E}_0e^{i(\bk\br-\omega t)}$ and I used equation (\ref{deltaGK}). After I insert these expressions into equation for the current, there will be several contributions, but I single out two contributions which contain the following combination of ${\mathbf E}$ and ${\mathbf E}^*$:
\beg\label{jdc}
\begin{split}
{\mathbf j}_{\textrm{dc}}(\omega)&=ie^2v_F^4\gamma_\omega\langle\bn(\bn{\mathbf E})(\bn{\mbox{\boldmath $\nabla$}})(\bn{\mathbf E}^*)\rangle_\bn\\&
-ie^2v_F^4\gamma_\omega\langle\bn(\bn{\mathbf E}^*)(\bn{\mbox{\boldmath $\nabla$}})(\bn{\mathbf E})\rangle_\bn.
\end{split}
\en
Alternatively, one can derive (\ref{jdc}) by averaging the full expression for current over a period of oscillations $2\pi/\omega$.
Clearly, upon averaging over the Fermi surface there is a contribution to the current $\propto{\mbox{\boldmath $\nabla$}}\times({\mathbf E}\times{\mathbf E}^*)$ which can be represented as 
${\mbox{\boldmath $\nabla$}}\times{\mathbf M}_{\textrm{ind}}$. Function $\gamma_\omega$ in expression (\ref{jdc}) is formally given by the following expression
\beg\label{gammaw}
\begin{split}
\gamma_\omega&=\frac{\nu_F^2}{8\omega}\int\limits_{-\infty}^\infty{\cal G}_\veps^K(\omega)d\veps
\int\limits_{-\infty}^\infty{\cal G}_{\veps'}^K(-\omega)d\veps'\equiv\frac{I(\omega,\tau)}{\omega^3}.
\end{split}
\en
Here on the last step I took into account that coefficients $a_\veps(\omega)$ and $b_\veps(\omega)$ are proportional to $1/\omega$ and, as a consequence, ${\cal G}_\veps^K(\omega)\propto1/\omega$, so that function $\gamma_\omega\propto 1/\omega^3$, while function $I(\omega,\tau)$ encodes the effects of disorder on the induced magnetization.

\begin{figure}
\includegraphics[width=0.935\linewidth]{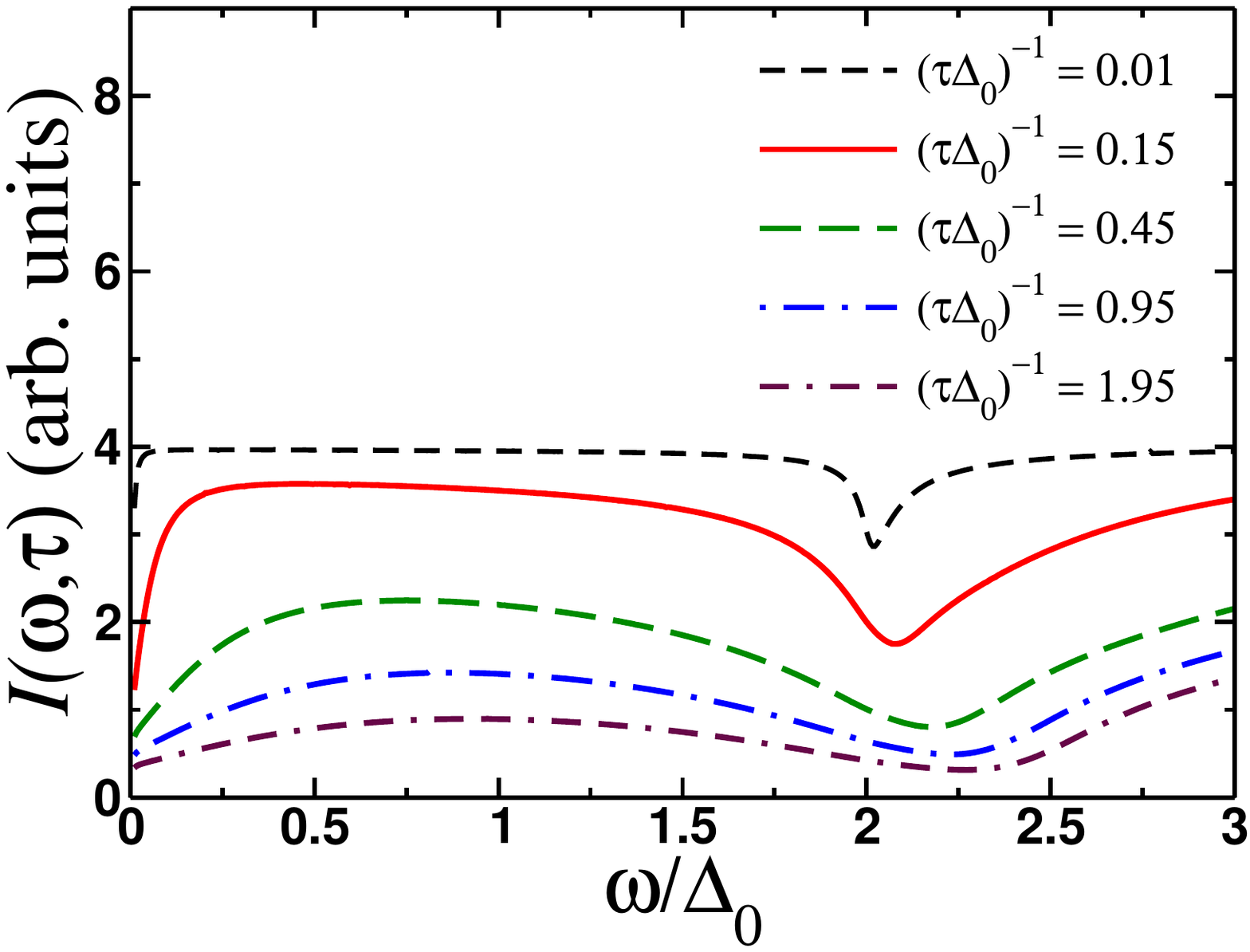}
\caption{Dependence of the function $I(\omega,\tau)$, Eq. (\ref{gammaw}), on frequency of external electric field for various values of the disorder scattering rate $\tau^{-1}$. Since function $I(\omega,\tau)$ determines the magnitude of the effect, from these results I conclude that disorder suppresses the value of the induced static magnetization. It is worthwhile to note that the suppression is strongest at frequencies $\omega_{\textrm{min}}\approx 2\Delta_0$, which coincides with the resonant frequency of the amplitude Higgs mode. The value of the temperature has been set to $T=0.1\Delta_0$.}
\label{Fig1}
\end{figure}

\begin{figure}
\includegraphics[width=0.935\linewidth]{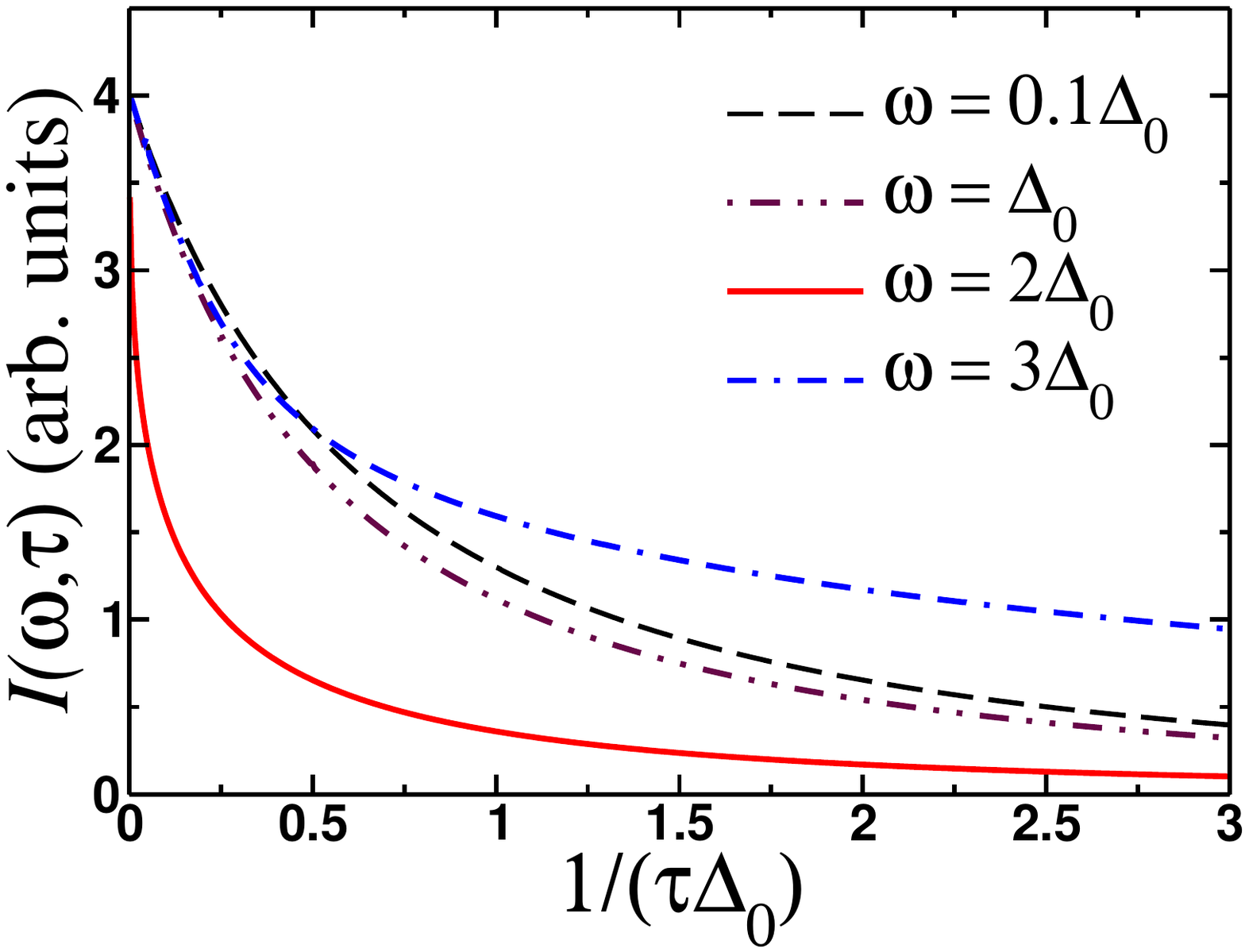}
\caption{Dependence of the function $I(\omega,\tau)$, Eq. (\ref{gammaw}), on disorder scattering rate for various values of frequency of an external electric field $\omega$. Note that suppression is much stronger at the resonant frequency of the amplitude Higgs mode $\omega=2\Delta_0$. The value of the temperature has been set to $T=0.1\Delta_0$.}
\label{Fig2}
\end{figure}

Function $I(\omega,\tau)$ appearing in (\ref{gammaw}) can be evaluated numerically for different values of the disorder scattering rate and temperature. In the dirty limit $\tau\Delta_0\ll 1$ this function will become weakly dependent on the disorder scattering rate.
In order to see this, I first note that in this limit both numerator and denominator in the expression for ${\cal G}_\veps^K$ will be varying linearly with $\tau^{-1}$. This is  because the terms which are proportional to $\tau^{-2}$ in denominator cancel out due to normalization condition (see Appendix B). In passing I note that this fact - the dependence of the Keldysh Green function on disorder scattering rate - is yet another manifestation of the observation made earlier that Anderson theorem does not apply to disordered nonequilibrium superconductors, i.e. and scattering on potential impurities does affect various out-of-equilibrium response functions, which conceptually is in agreement with the earlier results \cite{Silaev2019-Disorder,Yang2022-Disorder,Dzero2023-Disorder}.

In Fig. \ref{Fig1} I show the result of the calculation of the frequency dependence of function $I(\omega,\tau)$ for different values of 
$1/\tau\Delta_0$. As I can see this dependence is nonmonotonic. However, this function clearly acquires smaller values with an increase in $\tau^{-1}$. Note that $I(\omega,\tau)$ has a minimum at frequencies $\simeq 2\Delta_0$ which matches the value of the resonant frequency for the excitation of the amplitude Higgs mode. Notably, the minimum moves to higher frequencies with increase in the values of $\tau^{-1}$, which seems to be in agreement with the results of Ref.  \cite{Dzero2023-Disorder}. Since such a shift is not expected to appear in the strongly disordered superconductors \cite{Moore2017,Yantao2023}, I attribute this effect to the perturbative nature of the self-consistent Born approximation. 

In Fig. \ref{Fig2} the dependence of $I(\omega,\tau)$ on scattering rate $\tau^{-1}$ is shown. This dependence turns out to be perfectly monotonic and reflects the suppression of ${\mathbf M}_{\textrm{ind}}$ for four different values of $\omega$. Notably, the suppressions is by far the strongest when $\omega=2\Delta_0$, which is consistent with the results presented in Fig. \ref{Fig1}. I am inclined to interpret this suppression as being caused by the resonant excitation of the amplitude Higgs mode. 

\section{Conclusions}
In this paper using the quasiclassical theory of superconductivity I have investigated how potential impurities may affect the emergence of the inverse Faraday effect in conventional superconductors. Scattering on potential impurities have been taken into account within the self-consistent Born approximation. The normal contribution to the current must still be determined by the diffusion coefficient $D\propto\tau$, while the superfluid component of the current is suppressed by the pair breaking processes due to non-equilibrium \cite{Dzero2023-Disorder}. In agreement these qualitative observations, I have found that in the fairly broad range of frequencies potential impurities suppress the magnitude of the IFE. In addition, it has been found that the static magnetization ${\mathbf M}_{\textrm{ind}}(\omega)$ has a minimum for $\omega\approx 2\Delta_0$. This frequency also corresponds to the minimal excitation frequency for the amplitude Higgs mode. This observation provides another example of manifestation of the amplitude Higgs mode in response functions of a superconductor driven out-of-equilibrium \cite{Manske2014,Aoki2015,Kemper2015,Cea2016,Moore2017,Eremin2023,Yantao2023} and serves as a characteristic signature of the IFE in conventional superconductors. 

Apart from the quantitative difference between the results presented here and those of Ref. \cite{Balatsky2023} in which disorder effects were considered on the phenomenological level, I would like to emphasize that the microscopic description of the scattering on potential impurities in out-of-equilibrium superconductors is important for the proper account of the physics associated with the excitation of the amplitude Higgs mode. In this regard my results are in agreement with those of Ref. \cite{Silaev2019-Disorder} where it was shown how scattering on potential impurities leads to the effective excitation of the amplitude mode. 

I certainly hope that my theoretical prediction of the minimum in the dc magnetization as a function of frequency can be verified experimentally. Performing such a measurement could provide yet another important insight into the physics of the collective excitations and their impact on the IFE in conventional superconductors.  

\section{Acknowledgments}
I would like to thank A. V. Balatsky for bringing this problem to my attention and useful discussion. I also would like to thank Yantao Li for bringing several important references related to the main subject of this work to my attention. 
This work was financially supported by the National Science Foundation grant NSF-DMR-2002795. 

\begin{appendix}
\section{Auxiliary expressions}
In this Appendix I provide some auxiliary expressions which are used to derive the quasiclassical equation (\ref{Eq4g1RA}) in the main text. 
\paragraph{Time derivatives.} First simplification arises due to our choice for the time dependence of the electric field:
\beg\label{aux1}
\begin{split}
&\left[\partial_\veps\left(\check{g}_\veps\check{\tau}_3\right){\cal E}\left(\frac{i}{2}\stackrel{\leftarrow}\partial_\veps\stackrel{\rightarrow}\partial_t\right)+\partial_\veps\left(\check{\tau}_3\check{g}_\veps\right){\cal E}\left(-\frac{i}{2}\stackrel{\leftarrow}\partial_\veps\stackrel{\rightarrow}\partial_t\right)\right]e^{-i\omega t}\\&=\frac{e^{-i\omega t}}{2}\left[{\cal E}\left(-\frac{\omega}{2}\partial_\veps\right)+{\cal E}\left(\frac{\omega}{2}\partial_\veps\right)\right]\left\{\check{\tau}_3,\partial_\veps\check{g}_\veps\right\}\\&+\frac{e^{-i\omega t}}{2}\left[{\cal E}\left(-\frac{\omega}{2}\partial_\veps\right)-{\cal E}\left(\frac{\omega}{2}\partial_\veps\right)\right]\left[\check{\tau}_3,\partial_\veps\check{g}_\veps\right].
\end{split}
\en
Here I use the bare propagators since this terms always enter in the combination with external electric field, which is assumed to be small. 
This expression can be further simplified if I take into account the definition of function ${\cal E}$. For example, for the combination in the second line in (\ref{aux1}) it obtains:
\beg\label{SimplifyMore}
\begin{split}
&\left[{\cal E}\left(-\frac{\omega}{2}\partial_\veps\right)+{\cal E}\left(\frac{\omega}{2}\partial_\veps\right)\right]\partial_\veps\check{g}_\veps=\int\limits_{-1}^{1}e^{\eta(\omega/2)\partial_\veps}\partial_\veps\check g_{\veps}d\eta\\&=\partial_\veps\int\limits_{-1}^{1}\check g_{\veps+\eta\frac{\omega}{2}}d\eta=\frac{2}{\omega}\left(\check{g}_{\veps+\omega/2}-\check{g}_{\veps-\omega/2}\right), \\
\end{split}
\en
Similarly, for the expression in the third line in (\ref{aux1}) I find
\beg\label{SimplifyMore2}
\begin{split}
&\left[{\cal E}\left(\frac{\omega}{2}\partial_\veps\right)-{\cal E}\left(-\frac{\omega}{2}\partial_\veps\right)\right]\partial_\veps\check{g}_\veps=\partial_\veps\left(
\int\limits_0^1-\int\limits_{-1}^0\right)\check{g}_{\veps+\eta\omega/2}d\eta\\&=\frac{2}{\omega}\left(\check{g}_{\veps+\omega/2}+\check{g}_{\veps-\omega/2}-2\check{g}_\veps\right).
\end{split}
\en
Lastly, I note that the last term in the right hand side of this expression is canceled out by the term which contains the vector potential in the main equation, Eq. (\ref{EilenFin}). Thus, collecting all these contributions together and taking into account that the first order correction to the superconducting order parameter vanishes after the averaging over the Fermi surface yields equation (\ref{RALinear}) in the main text. 
\paragraph{Self energy part.} For the commutator which involves the self energy part $\check{\sigma}=-\frac{i}{2\tau}\langle\check{g}\rangle_\bn$ after retaining the linear order terms in powers of electric field I have
\beg\label{ucomm}
\begin{split}
&[\check{\sigma}_{\textrm{u}}\stackrel{\circ},\check{g}_\veps+\check{g}_1]\approx-\frac{i}{2\tau}[\check{g}_\veps\stackrel{\circ},\check{g}_1]
-\frac{i}{2\tau}[\langle\check{g}_1\rangle_\bn\stackrel{\circ},\check{g}_\veps]\\&=\frac{i}{2\tau}\check{g}_1e^{-\frac{i}{2}\stackrel{\leftarrow}\partial_t\stackrel{\rightarrow}\partial_\veps}\check{g}_\veps-\frac{i}{2\tau}\check{g}_\veps e^{\frac{i}{2}\stackrel{\leftarrow}\partial_\veps\stackrel{\rightarrow}\partial_t}\check{g}_1+\frac{i}{2\tau}\check{g}_\veps e^{\frac{i}{2}\stackrel{\leftarrow}\partial_\veps\stackrel{\rightarrow}\partial_t}\langle\check{g}_1\rangle_\bn\\&-\frac{i}{2\tau}\langle\check{g}_1\rangle_\bn e^{-\frac{i}{2}\stackrel{\leftarrow}\partial_t\stackrel{\rightarrow}\partial_\veps}\check{g}_\veps=\frac{i}{2\tau}\left(\check{g}_1-\langle \check{g}_1\rangle_\bn\right)\check{g}_{\veps-\frac{\omega}{2}}\\&-
\frac{i}{2\tau}\check{g}_{\veps+\frac{\omega}{2}}\left(\check{g}_1-\langle\check{g}_1\rangle_\bn\right).
\end{split}
\en
Note that the averages over the Fermi surface are nonzero only for the contributions to $\check{g}_1$ which are directly proportional to the dot product $\bn\bk$ (see e.g. (\ref{g1Ansatz}) in the main text), which means that averaging both sides of this equation over the Fermi surface will give zero in this case. This fact is used in the derivation of the equation (\ref{RALinearAv}) in the main text.

\section{Expressions for the linear corrections to the retarded and advanced propagators}
Components of the matrix function $\hat{\cal G}_\veps(\omega)$, Eq. (\ref{GRA}), must satisfy the following system of linear equations
\beg\label{SystemOne}
\begin{split}
&(\tilde{\veps}_\omega-\tilde{\veps}_{-\omega}){\cal G}_\veps(\omega)-(\tilde{\Delta}_\omega-\tilde{\Delta}_{-\omega}){\cal F}_\veps(\omega)=a_\veps(\omega), \\
&(\tilde{\Delta}_\omega+\tilde{\Delta}_{-\omega}){\cal G}_\veps(\omega)-(\tilde{\veps}_\omega+\tilde{\veps}_{-\omega}){\cal F}_\veps(\omega)=-b_\veps(\omega).
\end{split}
\en
Here I have omitted the $R(A)$ superscripts for brevity. 
The solution of these equations is
\beg\label{FirstOrderSol}
\begin{split}
{\cal G}_\veps(\omega)&=\frac{a_\veps(\omega)(\tilde{\veps}_{\omega}+\tilde{\veps}_{-\omega})+b_\veps(\omega)(\tilde{\Delta}_{\omega}-\tilde{\Delta}_{-\omega})}{\tilde{\veps}_{\omega}^2-\tilde{\veps}_{-\omega}^2-\tilde{\Delta}_{\omega}^2+\tilde{\Delta}_{-\omega}^2}, \\
{\cal F}_\veps(\omega)&=\frac{b_\veps(\omega)(\tilde{\veps}_{\omega}-\tilde{\veps}_{-\omega})+a_\veps(\omega)(\tilde{\Delta}_{\omega}+\tilde{\Delta}_{-\omega})}{\tilde{\veps}_{\omega}^2-\tilde{\veps}_{-\omega}^2-\tilde{\Delta}_{\omega}^2+\tilde{\Delta}_{-\omega}^2}.
\end{split}
\en
Note that in the dirty limit $\tau\Delta_0\ll 1$, the disorder scattering rate drops out from these expressions. Specifically, in the limit when
$\tau^{-1}\gg\textrm{max}\{\omega,\veps\}$ I have
\beg\label{ZnamDirty}
\begin{split}
&\tilde{\veps}_{\omega}^2-\tilde{\veps}_{-\omega}^2-\tilde{\Delta}_{\omega}^2+\tilde{\Delta}_{-\omega}^2\approx\frac{i\omega}{\tau}(g_{\veps+\frac{\omega}{2}}+g_{\veps-\frac{\omega}{2}})\\&+\frac{i\veps}{\tau}(g_{\veps+\frac{\omega}{2}}-g_{\veps-\frac{\omega}{2}})-
\frac{i\Delta_0}{\tau}(f_{\veps+\frac{\omega}{2}}-f_{\veps-\frac{\omega}{2}})
\end{split}
\en       
and the terms $O(1/\tau^2)$ cancel out due to normalization condition $g_\veps^2-f_\veps^2=1$. Since the numerator in Eq. (\ref{FirstOrderSol}) is also linear in $1/\tau$ the final result becomes independent of $\tau^{-1}$. This result is actually consistent both with the Anderson theorem and with the ones from  non-linear $\sigma$ model where the exact averaging over disorder is performed \cite{Kamenev2009,Kamenev2011} and then nonlinear response is studied using the Usadel equation \cite{Eremin2023,Yantao2023}. However, as it is shown in the main text, the Keldysh components retain their dependence on $\tau^{-1}$ for these functions contain all the information above non-equilibrium state of a superconductor.

\end{appendix}


\begin{thebibliography}{10}

\bibitem{THz1}
J.~A. F\"{u}l\"{o}p, L.~P\'{a}lfalvi, G.~Alm\'{a}si, and J.~Hebling, ``Design
  of high-energy terahertz sources based on optical rectification,'' {\em Opt.
  Express}, vol.~18, pp.~12311--12327, Jun 2010.

\bibitem{THz2}
D.~N. Basov, R.~D. Averitt, D.~van~der Marel, M.~Dressel, and K.~Haule,
  ``Electrodynamics of correlated electron materials,'' {\em Rev. Mod. Phys.},
  vol.~83, pp.~471--541, Jun 2011.

\bibitem{Shimano2012}
R.~Matsunaga and R.~Shimano, ``Nonequilibrium BCS state dynamics induced by
  intense terahertz pulses in a superconducting NbN film,'' {\em Phys. Rev.
  Lett.}, vol.~109, p.~187002, 2012.

\bibitem{Shimano2013}
R.~Matsunaga, Y.~I. Hamada, K.~Makise, Y.~Uzawa, H.~Terai, Z.~Wang, and
  R.~Shimano, ``Higgs amplitude mode in the BCS superconductors
  ${\mathrm{Nb}}_{1\mathrm{\text{-}}x}{\mathrm{Ti}}_{x}\mathrm{N}$ induced by
  terahertz pulse excitation,'' {\em Phys. Rev. Lett.}, vol.~111, p.~057002,
  Jul 2013.

\bibitem{Shimano2014}
R.~Matsunaga, N.~Tsuji, H.~Fujita, A.~Sugioka, K.~Makise, Y.~Uzawa, H.~Terai,
  Z.~Wang, H.~Aoki, and R.~Shimano, ``Light-induced collective pseudospin
  precession resonating with Higgs mode in a superconductor,'' {\em Science},
  vol.~345, no.~6201, pp.~1145--1149, 2014.

\bibitem{THz3}
M.~Beck, I.~Rousseau, M.~Klammer, P.~Leiderer, M.~Mittendorff, S.~Winnerl,
  M.~Helm, G.~N. Gol'tsman, and J.~Demsar, ``Transient increase of the energy
  gap of superconducting NbN thin films excited by resonant narrow-band
  terahertz pulses,'' {\em Phys. Rev. Lett.}, vol.~110, p.~267003, Jun 2013.

\bibitem{Millis2020}
Z.~Sun, M.~M. Fogler, D.~N. Basov, and A.~J. Millis, ``Collective modes and
  terahertz near-field response of superconductors,'' {\em Phys. Rev. Res.},
  vol.~2, p.~023413, Jun 2020.

\bibitem{Armi2014}
Y.~Lubashevsky, L.~Pan, T.~Kirzhner, G.~Koren, and N.~P. Armitage, ``Optical
  birefringence and dichroism of cuprate superconductors in the THz regime,''
  {\em Phys. Rev. Lett.}, vol.~112, p.~147001, Apr 2014.

\bibitem{Armitage2023}
K.~Katsumi, J.~Fiore, M.~Udina, R.~R.~I. au2, D.~Barbalas, J.~Jesudasan,
  P.~Raychaudhuri, G.~Seibold, L.~Benfatto, and N.~P. Armitage, ``Revealing
  novel aspects of light-matter coupling in terahertz two-dimensional coherent
  spectroscopy: the case of the amplitude mode in superconductors,'' 2023.

\bibitem{Varma2014}
D.~Pekker and C.~Varma, ``Amplitude/Higgs modes in condensed matter physics,''
  {\em Annual Review of Condensed Matter Physics}, vol.~6, no.~1, pp.~269--297,
  2015.

\bibitem{Cea2016}
T.~Cea, C.~Castellani, and L.~Benfatto, ``Nonlinear optical effects and
  third-harmonic generation in superconductors: Cooper pairs versus Higgs mode
  contribution,'' {\em Phys. Rev. B}, vol.~93, p.~180507, May 2016.

\bibitem{Shimano2020}
R.~Shimano and N.~Tsuji, ``Higgs mode in superconductors,'' {\em Annual Review
  of Condensed Matter Physics}, vol.~11, no.~1, pp.~103--124, 2020.

\bibitem{Haenel2021-Disorder}
R.~Haenel, P.~Froese, D.~Manske, and L.~Schwarz, ``Time-resolved optical
  conductivity and Higgs oscillations in two-band dirty superconductors,'' {\em
  Phys. Rev. B}, vol.~104, p.~134504, Oct 2021.

\bibitem{Sherman2015-Disorder}
D.~Sherman, U.~S. Pracht, B.~Gorshunov, S.~Poran, J.~Jesudasan, M.~Chand,
  P.~Raychaudhuri, M.~Swanson, N.~Trivedi, A.~Auerbach, M.~Scheffler,
  A.~Frydman, and M.~Dressel, ``The Higgs mode in disordered superconductors
  close to a quantum phase transition,'' {\em Nature Physics}, vol.~11, no.~2,
  pp.~188--192, 2015.

\bibitem{Spivak-Higgs}
M.~Bellitti, C.~R. Laumann, and B.~Z. Spivak, ``Incoherent excitation of
  coherent Higgs oscillations in superconductors,'' {\em Phys. Rev. B},
  vol.~105, p.~104513, Mar 2022.

\bibitem{Eliashberg1970}
G.~Eliashberg, ``Film superconductivity stimulated by a high-frequency field,''
  {\em Sov. Phys. - JETP Lett.}, vol.~11, p.~114, 1970.

\bibitem{Eremin2023}
P.~Derendorf, A.~F. Volkov, and I.~Eremin, ``Nonlinear response of diffusive
  superconductors to ac-electromagnetic fields,'' {\em arXiv:2308.00838}, 2023.

\bibitem{Chen2017}
X.-J. Chen, ``Fundamental mechanism for all-optical helicity-dependent
  switching of magnetization,'' {\em Scientific Reports}, vol.~7, no.~1,
  p.~41294, 2017.

\bibitem{Croitoru2022}
M.~D. Croitoru, S.~V. Mironov, B.~Lounis, and A.~I. Buzdin, ``Toward the
  light-operated superconducting devices: Circularly polarized radiation
  manipulates the current-carrying states in superconducting rings,'' {\em
  Advanced Quantum Technologies}, vol.~5, no.~10, p.~2200054, 2022.

\bibitem{SpivakGiant2020}
M.~Smith, A.~V. Andreev, and B.~Z. Spivak, ``Debye mechanism of giant microwave
  absorption in superconductors,'' {\em Phys. Rev. B}, vol.~101, p.~134508, Apr
  2020.

\bibitem{Pit1961}
L.~P. Pitaevskii, ``Electric forces in a transparent dispersive medium,'' {\em
  Sov. Phys. - JETP}, vol.~12, p.~1008, 1961.

\bibitem{Pershan1966}
P.~S. Pershan, J.~P. van~der Ziel, and L.~D. Malmstrom, ``Theoretical
  discussion of the inverse Faraday effect, Raman scattering, and related
  phenomena,'' {\em Phys. Rev.}, vol.~143, pp.~574--583, Mar 1966.

\bibitem{Battiato2014}
M.~Battiato, G.~Barbalinardo, and P.~M. Oppeneer, ``Quantum theory of the
  inverse Faraday effect,'' {\em Phys. Rev. B}, vol.~89, p.~014413, Jan 2014.

\bibitem{yang2022inverse}
X.~Yang, Y.~Mou, H.~Zapata, B.~Reynier, B.~Gallas, and M.~Mivelle, ``An inverse
  Faraday effect through linear polarized light,'' 2022.

\bibitem{mou2023reversed}
Y.~Mou, X.~Yang, B.~Gallas, and M.~Mivelle, ``A reversed inverse Faraday
  effect,'' 2023.

\bibitem{mou2023chiral}
Y.~Mou, X.~Yang, B.~Gallas, and M.~Mivelle, ``A chiral inverse Faraday effect
  mediated by an inversely designed plasmonic antenna,'' 2023.

\bibitem{IFE-Mott2022}
S.~Banerjee, U.~Kumar, and S.-Z. Lin, ``Inverse Faraday effect in Mott
  insulators,'' {\em Phys. Rev. B}, vol.~105, p.~L180414, May 2022.

\bibitem{gao2020topological}
Y.~Gao, C.~Wang, and D.~Xiao, ``Topological inverse Faraday effect in 
Weyl semimetals,'' 2020.

\bibitem{parchenko2023plasmonenhanced}
S.~Parchenko, K.~Hofhuis, A.~Ciuciulkaite, V.~Kapaklis, V.~Scagnoli,
  L.~Heyderman, and A.~Kleibert, ``Plasmon-enhanced optical control of
  magnetism at the nanoscale via the inverse Faraday effect,'' 2023.

\bibitem{Han_2023}
J.~W. Han, P.~Sai, D.~B. But, E.~Uykur, S.~Winnerl, G.~Kumar, M.~L. Chin, R.~L.
  Myers-Ward, M.~T. Dejarld, K.~M. Daniels, T.~E. Murphy, W.~Knap, and
  M.~Mittendorff, ``Strong transient magnetic fields induced by THz-driven
  plasmons in graphene disks,'' {\em Nature Communications}, vol.~14, Nov.
  2023.

\bibitem{Majedi2021}
A.~H. Majedi, ``Microwave-induced inverse Faraday effect in superconductors,''
  {\em Phys. Rev. Lett.}, vol.~127, p.~187001, 2021.

\bibitem{Mironov2021-IFESC}
S.~V. Mironov, A.~S. Mel'nikov, I.~D. Tokman, V.~Vadimov, B.~Lounis, and A.~I.
  Buzdin, ``Inverse Faraday effect for superconducting condensates,'' {\em
  Phys. Rev. Lett.}, vol.~126, p.~137002, Apr 2021.

\bibitem{Parafilo2022Fl}
A.~V. Parafilo, M.~V. Boev, V.~M. Kovalev, and I.~G. Savenko, ``Photogalvanic
  transport in fluctuating Ising superconductors,'' {\em Phys. Rev. B},
  vol.~106, p.~144502, Oct 2022.

\bibitem{Putilov2023-IFESC}
A.~V. Putilov, S.~V. Mironov, A.~S. Mel'nikov, and A.~A. Bespalov, ``Inverse
  Faraday effect in superconductors with a finite gap in the excitation
  spectrum,'' {\em JETP Letters}, vol.~117, no.~11, pp.~827--833, 2023.

\bibitem{Buzdin2023}
V.~Plastovets and A.~Buzdin, ``Fluctuation-mediated inverse Faraday effect in
  superconducting rings,'' {\em Physics Letters A}, vol.~481, p.~129001, 2023.

\bibitem{Croitoru2023}
M.~D. Croitoru, B.~Lounis, and A.~I. Buzdin, ``{Helicity-controlled switching
  of superconducting states by radiation pulse},'' {\em Applied Physics
  Letters}, vol.~123, p.~122601, 09 2023.

\bibitem{Balatsky2023}
P.~Sharma and A.~V. Balatsky, ``Light induced magnetism in metals via inverse
  Faraday effect,'' 2023.

\bibitem{AndersonTheorem}
P.~W. Anderson, ``Knight shift in superconductors,'' {\em Phys. Rev. Lett.},
  vol.~3, pp.~325--326, Oct 1959.

\bibitem{AG1961}
A.~A. Abrikosov and L.~P. Gor'kov, ``Contribution to the theory of
  superconducting alloys with paramagnetic impurities,'' {\em Sov. Phys. -
  JETP}, vol.~12, p.~1243, 1961.

\bibitem{Balatsky-RMP}
A.~V. Balatsky, I.~Vekhter, and J.-X. Zhu, ``Impurity-induced states in
  conventional and unconventional superconductors,'' {\em Rev. Mod. Phys.},
  vol.~78, pp.~373--433, May 2006.

\bibitem{Silaev2019-Disorder}
M.~Silaev, ``Nonlinear electromagnetic response and Higgs-mode excitation in
  BCS superconductors with impurities,'' {\em Phys. Rev. B}, vol.~99,
  p.~224511, Jun 2019.

\bibitem{Yang2022-Disorder}
F.~Yang and M.~W. Wu, ``Impurity scattering in superconductors revisited:
  Diagrammatic formulation of the supercurrent-supercurrent correlation and
  Higgs-mode damping,'' {\em Phys. Rev. B}, vol.~106, p.~144509, Oct 2022.

\bibitem{Yang2022-Disorder2}
F.~Yang and M.~W. Wu, ``Influence of scattering on the optical response of superconductors,'' {\em Phys. Rev. B}, vol.~102, p.~144508, Oct 2020.

\bibitem{Dzero2023-Disorder}
M.~Dzero, ``Collisionless dynamics in disordered superconductors,'' {\em
  arXiv:2303.06750}, 2023.

\bibitem{LO}
A.~I. Larkin and Y.~Ovchinnikov, ``Quasiclassical method in the theory of
  superconductivity,'' {\em Sov. Phys. - JETP}, vol.~28, p.~1200, 1969.

\bibitem{Usadel1970}
K.~D. Usadel, ``Generalized diffusion equation for superconducting alloys,''
  {\em Phys. Rev. Lett.}, vol.~25, pp.~507--509, Aug 1970.

\bibitem{Kita2001}
T.~Kita, ``Gauge invariance and Hall terms in the quasiclassical equations of
  superconductivity,'' {\em Phys. Rev. B}, vol.~64, p.~054503, Jun 2001.

\bibitem{Kopnin1994}
N.~B. Kopnin, ``Kinetic equations for clean superconductors: Application to the
  flux flow Hall effect,'' {\em Journal of Low Temperature Physics}, vol.~97,
  no.~1, pp.~157--179, 1994.

\bibitem{Kamenev2009}
A.~Kamenev and A.~Levchenko, ``Keldysh technique and non-linear $\sigma$-model:
  basic principles and applications,'' {\em Advances in Physics}, vol.~58,
  no.~3, pp.~197--319, 2009.

\bibitem{Kamenev2011}
A.~Kamenev, {\em Field Theory of Non-Equilibrium Systems}.
\newblock Cambridge University Press, 2011.

\bibitem{Kita-Review}
T.~Kita, ``{Introduction to Nonequilibrium Statistical Mechanics with Quantum
  Field Theory},'' {\em Progress of Theoretical Physics}, vol.~123,
  pp.~581--658, 04 2010.

\bibitem{Yantao2023}
Y.~Li and M.~Dzero, ``Amplitude Higgs mode in superconductors with magnetic
  impurities,'' {\em arXiv:2311.09310}, 2023.

\bibitem{Moore2017}
A.~Moor, A.~F. Volkov, and K.~B. Efetov, ``Amplitude Higgs mode and admittance
  in superconductors with a moving condensate,'' {\em Phys. Rev. Lett.},
  vol.~118, p.~047001, Jan 2017.

\bibitem{Manske2014}
H.~Krull, D.~Manske, G.~S. Uhrig, and A.~P. Schnyder, ``Signatures of
  nonadiabatic BCS state dynamics in pump-probe conductivity,'' {\em Phys. Rev.
  B}, vol.~90, p.~014515, Jul 2014.

\bibitem{Aoki2015}
N.~Tsuji and H.~Aoki, ``Theory of Anderson pseudospin resonance with Higgs mode
  in superconductors,'' {\em Phys. Rev. B}, vol.~92, p.~064508, Aug 2015.

\bibitem{Kemper2015}
A.~F. Kemper, M.~A. Sentef, B.~Moritz, J.~K. Freericks, and T.~P. Devereaux,
  ``Direct observation of Higgs mode oscillations in the pump-probe
  photoemission spectra of electron-phonon mediated superconductors,'' {\em
  Phys. Rev. B}, vol.~92, p.~224517, Dec 2015.

\end{thebibliography}

\end{document}